\documentstyle[12pt,blois,psfig]{article}
\begin{document}
\renewcommand{\author}[4]{
\vspace{5mm}
\begin{center}
        {\normalsize \rm #1}\\
        {\normalsize \it #2}\\
        {\normalsize \it #3}\\
        {\normalsize \it #4}\\
        \vspace{1cm}
\end{center}
}
\newcommand{\lya}{\mbox{${\rm Ly}\alpha$}}
\newcommand{\kms}{\mbox{km\ s${^{-1}}$}}
\def \apjl#1#2  {{\em Astrophys. J. Lett. \/} {\bf #1}, {#2}}
\def \araa#1#2  {{\em Ann. Rev. Astron. Astrophys. \/} {\bf #1}, {#2}}
\def\lsim{\mathrel{\mathchoice {\vcenter{\offinterlineskip\halign{\hfil
$\displaystyle##$\hfil\cr<\cr\sim\cr}}}
{\vcenter{\offinterlineskip\halign{\hfil$\textstyle##$\hfil\cr
<\cr\sim\cr}}}
{\vcenter{\offinterlineskip\halign{\hfil$\scriptstyle##$\hfil\cr
<\cr\sim\cr}}}
{\vcenter{\offinterlineskip\halign{\hfil$\scriptscriptstyle##$\hfil\cr
<\cr\sim\cr}}}}}
\def\gsim{\mathrel{\mathchoice {\vcenter{\offinterlineskip\halign{\hfil
$\displaystyle##$\hfil\cr>\cr\sim\cr}}}
{\vcenter{\offinterlineskip\halign{\hfil$\textstyle##$\hfil\cr
>\cr\sim\cr}}}
{\vcenter{\offinterlineskip\halign{\hfil$\scriptstyle##$\hfil\cr
>\cr\sim\cr}}}
{\vcenter{\offinterlineskip\halign{\hfil$\scriptscriptstyle##$\hfil\cr
>\cr\sim\cr}}}}}
%
% -----------------------------------------------------------------------------
%
\heading{ENVIRONMENTAL DEPENDENCE OF THE NATURE OF GASEOUS
ENVELOPES OF GALAXIES at $z<1$}

\author{N. Yahata$^{1}$, K. M. Lanzetta$^{1}$, J. K. Webb$^{2}$, \& X. Barcons$^{3}$}
{$^{1}$State University of New York at Stony Brook, Stony Brook, U.S.A.}
{$^{2}$University of New South Wales, Sydney, Australia}
{$^{3}$Instituto de F{\'\i}sica de Cantabria, Universidad de Cantabria, Santander, Spain}

\begin{bloisabstract}
We present new statistical study on the origin of \lya\ absorption systems
that are detected in QSO fields.  The primary objective is to investigate
environmental dependence of the relationship between galaxies and \lya\ absorption
systems.  We find that galaxies exhibit no environmental dependence in giving rise
to \lya\ absorbers in their halos.  Conversely, \lya\ absorbers do not exhibit
preference as to the environment where they reside.
\end{bloisabstract}

\section{Introduction}
Understanding the environmental dependence of the relationship between
galaxies and \lya\ absorption systems is crucial to understanding the nature of
extended gas around galaxies and the origin of \lya\ absorption systems.  Early
observations of high-redshift \lya\ absorption systems indicated that the
absorbers are unclustered or only weakly clustered in redshift, suggesting that
the absorbers are not distributed like galaxies \cite{Sargent80}.
More recent observations of the large-scale galaxy environments of low-redshift
\lya\ absorption systems indicate that the absorbers are less clustered with
galaxies than galaxies are clustered with each other, suggesting that the
absorbers trace the same large-scale structures as galaxies but are not
directly associated with individual galaxies
\cite{Morris93}\cite{Bowen96}\cite{LeBrun96}.

Yet recent observations by Lanzetta et al.\ (1995) indicate that many or most
\lya\ absorption systems at redshifts $z < 1$ arise in extended gaseous
envelopes of galaxies of roughly $160 \ h^{-1}$ kpc radius and unit covering
factor, suggesting that the absorbers are directly associated with individual
galaxies \cite{Lanzetta95}\cite{Chen98}. There is also an identification of two
groups or clusters of galaxies that do produce corresponding \lya\ absorption
lines, demonstrating that at least some groups or clusters of galaxies produce
\lya\ absorption systems and that at least some \lya\ absorption systems arise
in groups or clusters of galaxies \cite{Lanzetta97}\cite{OrtizGil98}.

If the absorbers are directly associated with individual galaxies, then
why is it that the absorbers are apparently unclustered or only weakly clustered
in redshift and are apparently less clustered with galaxies than galaxies are
clustered with each other?  These results may be reconciled if (1) the
absorbers generally arise in galaxies but avoid galaxies in regions of higher
than average density and are thus significantly less clustered than galaxies
or (2) the clustering measurements underestimate the actual clustering of the
absorbers.  To address these issues, it is necessary to establish the environmental
dependence of the relationship between galaxies and absorption systems.  Here we
investigate this relationship at redshifts $z < 1$ based on an ongoing imaging
and spectroscopic survey of faint galaxies in fields of Hubble Space Telescope
(HST) spectroscopic target QSOs
\cite{Lanzetta95}\cite{Chen98}\cite{OrtizGil98}\cite{Barcons95}\cite{Lanzetta96}.

\section{Galaxy and Absorber Data}
The analysis is based upon an ongoing imaging and spectroscopic survey of
faint galaxies in fields of HST spectroscopic target QSOs.  Detailed
descriptions of the observations have been and will be presented elsewhere
(e.g. \cite{Lanzetta95}), but in summary the observations consist of (1)
optical images and spectroscopy of objects in the fields of the QSOs, obtained
with various telescopes and from the literature, and (2) ultraviolet
spectroscopy of the QSOs, obtained with HST using the Faint Object Spectrograph
(FOS) and accessed through the HST archive.

The current sample contains 348 galaxies and 229 absorbers toward 24 QSOs. 
Redshifts of the galaxies span $z_{\rm gal} = 0.0218 - 0.7873$, with a median of
$z = 0.2998$.  Redshifts of the absorbers span $z_{\rm abs} = 0.0155 - 1.0361$,
with a median of $z = 0.4679$.  Absorption systems are detected according to a
$5 \sigma$ detection threshold.  Measurement uncertainties in redshift are
typically $\approx 150$ \kms\ for galaxy redshifts and $\approx 30$ \kms\ for
absorber redshifts.  Because the object of the present analysis is to examine
the correlations of intervening (rather than intrinsic) galaxies and absorbers,
it is necessary to exclude galaxies and absorbers that are physically related to
the QSOs.  Accordingly, we exclude from the analysis all galaxies and absorbers
with relative velocities to the background QSOs within $\pm 3000$ \kms.  This
leaves 263 galaxies and 212 absorbers toward 24 QSOs included into the analysis.

\section{Environment Traced through Galaxy Auto-Correlation Functions}
Because galaxies of the sample are sparsely sampled, it is impossible for us to
establish the local galaxy environment around any particular absorber to within
a meaningful degree of precision.  Instead, we must apply a {\em statistical}
measure of the local galaxy environment.  It is well known that galaxies in the
local universe exhibit a morphology--density relationship in that early-type
galaxies are preferentially found in regions of higher-than-average galaxy
density while late-type galaxies are preferentially found in regions of
lower-than-average galaxy density.  This relationship applies from galaxy
clusters through sparse groups through the field (see \cite{Dressler84} and
references therein).  The existence of a morphology--density relationship implies
that early-type galaxies typically trace regions of higher-than-average galaxy
density while late-type galaxies typically trace region of lower-than-average galaxy
density---in other words, galaxy morphology is a measure of local galaxy environment
in a statistical sense.

We use measurements of spectroscopic morphology to divide the galaxy sample
into early-type and late-type samples.  Figure 1(a) shows the galaxy auto-correlation
function for early-type (43 galaxies) and late-type galaxies (220 galaxies) at
impact parameter range $0\leq\rho\leq200$\ $h_0^{-1}$\ kpc.  It is seen that
early-type galaxies are more strongly clustered (by a factor of $10.0 \pm 2.7$
in terms of the correlation amplitude within $v = 1000$\ km s$^{-1}$), confirming
that the morphology--density relationship established in the local universe also
applies at higher redshifts ($z_{\rm med}\approx0.3$).

\section{Environmental Dependence}
The galaxy--absorber cross-correlation function (GACCF), $\xi_{\rm ga}(v,\rho)$,
is defined in such a way that the probability of finding an absorber within $dv$
of a randomly chosen galaxy is
\begin{equation}
dP = m_a H_0^{-1} [ 1 + \xi_{\rm ga}(v,\rho)] dv,
\end{equation}
where $m_a$ is the absorber number density per unit length and $H_0$ is the Hubble
constant.  Figure 1(b) shows the GACCF for early- and late-type galaxies in our sample.
We find that they are statistically identical to each other, which implies that the
absorbers do not strongly favor or avoid regions of higher-than-average or
lower-than-average galaxy density, at least over the range of densities probed by
the current samples.

The GACCF is dominated by different functions according to the scale in impact
parameter (see \cite{Yahata98} for detailed discussion).  First, at small impact
parameters ($\rho\lsim\rho_0\approx 200$\ kpc, \cite{Lanzetta95}\cite{Chen98}), the
GACCF is dominated by the galaxy halo distribution function.  Integrating the GACCF
within $\rho\lsim\rho_0$ then yields the ``integrated probability'' that a given
galaxy produces a \lya\ absorption system somewhere in its halo.  For our sample,
we find that this integral does not vary substantially according to galaxy
type---the ratio of the integral for early-type to late-type galaxies is $1.6\pm0.7$.
Together with the morphology--density relation established in the sample, this
suggests that {\em there is no significant environmental dependence for galaxy
halos that produce \lya\ absorption systems.}

Second, at large impact parameters ($\rho\gsim\rho_0$), the GACCF is dominated by
galaxy correlation functions, specifically:
\begin{eqnarray}
\xi_{\rm 1a} & = & f_1\xi_1 + f_2\xi'_{21}\\
\xi_{\rm 2a} & = & f_2\xi_2 + f_1\xi'_{12}\ ,
\end{eqnarray}
where $\xi_{\rm ia}$ is the GACCF, $\xi_i$ is the galaxy auto-correlation function,
$\xi'_{ij}$ is the galaxy--galaxy cross-correlation function ($i\left(\neq j\right)=
\left\{1,2\right\}$ specify galaxy type; 1 for early-type, 2 for late-type).  Here,
$f_1, f_2$ are the fractions of \lya\ absorbers that arise in halos of early- and
late-type galaxies, respectively, which from our sample we find $f_1=0.07\pm0.12$
and $f_2=0.59\pm0.16$.  Note that the ratio $\left(f_2/f_1\approx 8\right)$ is roughly
consistent with the ratio of early- to late-type galaxies in the sample.  Therefore,
these results suggest that {\it the \lya\ absorbers arise in halos of galaxies
irrespective of the galaxy type, and hence the environment where they reside.}

\section{Conclusion}
We investigated the environmental dependence of the relationship between galaxies and
absorption systems.  We confirmed that the morphology--density is established in our
galaxy sample ($z_{\rm med}\approx0.3$) in that early-type galaxies are more strongly
clustered than late-type galaxies.  Within this environment we found that galaxies
exhibit no environmental dependence in giving rise to \lya\ absorbers in their halos.
Conversely, \lya\ absorbers do not exhibit preference as to the environment where
they reside.  We therefore conclude that the previous clustering measurements of
\lya\ absorbers is likely to have underestimated their actual clustering properties.

\acknowledgements{This research was supported by NASA grant NAGW--4422 and NSF
grant AST--9624216.}

\begin{figure}
\centerline{\psfig{file=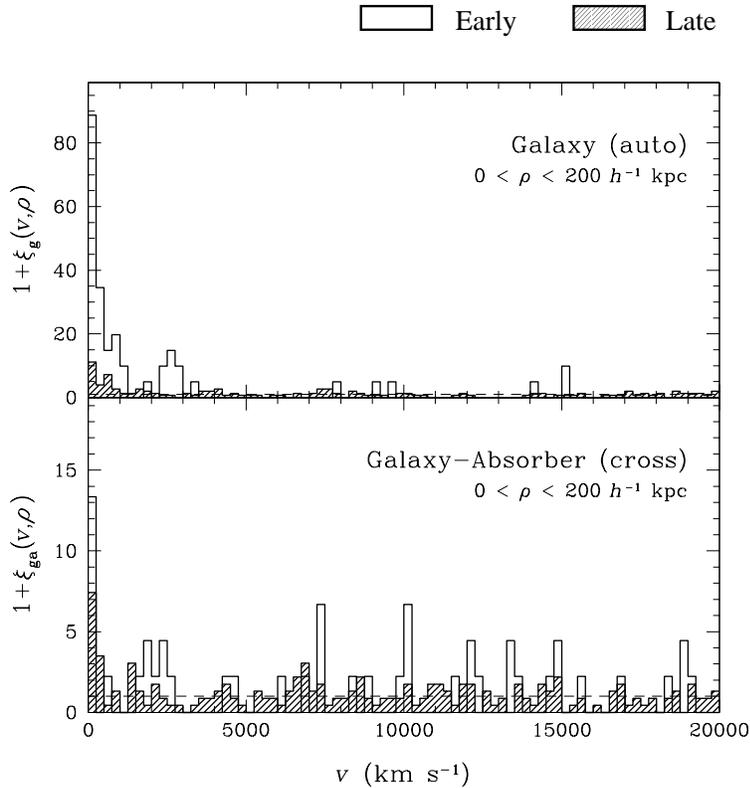,width=4.0in}}
\caption{
(a) [top] Galaxy auto-correlation functions for early- and late-type galaxies.
(b) [bottom] Galaxy--absorber cross-correlation functions.
}
\end{figure}

\vfill
\end{document}